\documentclass[submission]{eptcs}

\usepackage{personal}
\usepackage{stmaryrd}
\usepackage{wrapfig}
\usepackage{cite}
\usepackage{textcomp}

\author{Philippa Cowderoy \email{flippa@flippac.org}}
\title{Information Aware Type Systems and \\ Telescopic Constraint Trees}

\date{27th March 2020}

\begin{document}
\maketitle

\begin{abstract}
Can we use the flow of information to understand type systems? I present two familiar type systems in pursuit of an `Information Aware' style, using information effects to reveal data flow and help in implementing them. I also calculate a general, scoped, constraint-based representation of typechecking problems from the typing rules.
\end{abstract}

\section{Introduction and Background}

How does information arise and flow in type systems?

I will attempt to answer this by presenting type systems in an `Information Aware' style, making heavy use of information effects\cite{DBLP:conf/popl/JamesS12} which first arose in reversible programming. The paper introduces a combinator language for `isomorphic programming' in the same sense as functional programming. Information effects show where, in an otherwise isomorphic program, information is created or destroyed.

In an interpreter or semantics, the created information is the meaning of the input program. Correspondingly, in a type system information effects help to keep track of meaning and how it relates to the flow of information. By following the origin and flow of information we can derive algorithms, gain intuitions about algorithmic behaviour, and abstract away implementation details. Here I apply this to type checking and inference.


I also use constraints as a critical abstraction. Firstly they act as a unit of information. They also have both static and dynamic semantics: satisfaction predicates state when a constraint is satisfied, separately from when or how constraints can be solved.

Describing type systems using constraints is constraint logic programming. Such programs often won't correspond to a single isomorphic program -- instead, procedures implementing them for specific modes do\cite{Reddy1993}\cite{Rohwedder1996}. Typings of known source terms are programs, generating constraint problems.

\section{Information Aware Simply Typed Lambda Calculus}
\label{sec:IASTLC}

\begin{wrapfigure}{r}{0pt}
\begin{tabular}{r l}
$\mty \eqCon \mty$&Type equality\\
$\mty\dupCon^{\mty{}}_{\mty{}}$&Type duplication\\\\
$x:\mty \inCon \ctxt$&Binding in context\\
$\ctxt \assignCon \ctxt \;\consCon x:\mty$&Context extension\\
$\ctxt\dupCon^{\ctxt{}}_{\ctxt{}}$&Context duplication\\
\end{tabular}
\caption{Constraints for STLC}
\label{fig:STLCConstraints}
\end{wrapfigure}

The type system in figure \ref{fig:IASTLC} uses constraints shown in figure \ref{fig:STLCConstraints}. The system itself is a presentation of the conventional simply typed lambda calculus without type annotations. 

Due to the discipline imposed by information effects, metavariables are linear: used at one source and one sink. Duplication must be explicit as it creates copies of information. Constraint generation also creates information, as does constructing the function arrow -- something which we are inferring from the source.

\clearpage 

To truly explain the semantics of a typed lambda calculus, we need not just the types but the semantics of the context -- including the structural rules, encoded into the semantics of the context constraints. These can even be altered to produce a minimal substructural system.

The system supports any modes a solver is available for -- though proof search remains hard! Examples can be found in appendix \ref{app:modes}. Supporting annotations requires type duplication -- see appendix \ref{app:annotation} for details.

\begin{figure}[h]
\begin{mathpar}
\inferrule*{x : \mty \inCon \ctxt}
{\ctxt \turnstile x : \mty}
{Var}

\inferrule*{\ctxt{f} \assignCon \ctxt \; \consCon x : \mty{p}\\\\
\ctxt{f} \turnstile T : \mty{r}\\\\
\mty{f} \eqCon \mty{p} \;\funcr\; \mty{r}}
{\ctxt \turnstile \lambda x . T \; : \; \mty{f}}
{Lam}

\inferrule*{\ctxt{}\dupCon^{\ctxt{f}}_{\ctxt{p}}\\\\
\ctxt{f} \turnstile {T\!f} : \mty{f}\\
\ctxt{p} \turnstile {T\!p} : \mty{p}\\\\
\mty{p} \;\funcr\; \mty{r} \eqCon \mty{f}}
{\ctxt{} \turnstile {T\!f} \;\;\; {T\!p} \;:\; \mty{r}}
{App}
\end{mathpar}
\caption{Information Aware Simply Typed Lambda Calculus}
\label{fig:IASTLC}
\end{figure}

The $Var$ rule asserts that the binding $x:\mty{}$ is in the context $\ctxt{}$. $Lam$ uses a context extension to relate the context of the lambda to that of its body, with an equality constraint spelling out the relationship between the parameter type, the body type and the function represented by the lambda. $App$ duplicates its context in order to type its subterms and the equality constraint relates function, parameter and result.

Below is a mode analysis for type system used at the mode $\ctxt^+ \turnstile T^+ : \mty^-$, aka type synthesis or `typechecking'. It's a worthwhile exercise to connect each +ve variable to its -ve counterpart: constraints can fire when all their -ve inputs are available, producing +ve outputs (possibly unknown!) and showing the flow of data.

\begin{mathpar}
\inferrule*[vskip=1ex]{x^- : \mty^+ \inCon \ctxt^-}
{\ctxt^+ \turnstile x^+ : \mty^-}
{Var}

\inferrule*[vskip=1ex]{\ctxt{f}^+ \assignCon \ctxt^- \; \consCon x^- : \mty{p}^+\\\\
\ctxt{f}^- \turnstile T^- : \mty{r}^+\\\\
\mty{f}^+ \eqCon \mty{p}^- \;\funcr^+\; \mty{r}^-}
{\ctxt^+ \turnstile \lambda x^+ . T^+ \; : \; \mty{f}^-}
{Lam}

\inferrule*[vskip=1ex]{\ctxt^-\dupCon^{\ctxt{f}^+}_{\ctxt{p}^+}\\\\
\ctxt{f}^- \turnstile Tf^- : \mty{f}^+\\
\ctxt{p}^- \turnstile Tp^- : \mty{p}^+\\\\
\mty{p}^- \;\funcr^-\; \mty{r}^+ \eqCon \mty{f}^-}
{\ctxt^+ \turnstile {T\!f}^+ \; {T\!p}^+ \;:\; \mty{r}^-}
{App}
\end{mathpar}

\section{Telescopic Constraint Trees}

\begin{wrapfigure}{r}{0pt}
\begin{tabular}{r|l l}
$\exCon\mty$&Bind an unknown $\mty$\\
$\exConIs{\mty}\mty$&Bind $\mty$ with current solution\\
$\bask{x:\mty}$&Query/Ask for current binding [here]\\
$\btell{x:\mty}$&Generate/Tell about binding [here]
\end{tabular}
\caption{Telescopic constraints}
\label{fig:TelescopicConstraints}
\end{wrapfigure}

A telescopic constraint tree is a refinement of the AST into a de Bruijn telescope extended with branching (represented here using $|$) and scoped constraint machinery. The tree traces in space where a traditional checker will go over time. This provides a general representation of typechecking problems.

Figure \ref{fig:TelescopicSTLC} contains a CPS-like transformation using the quantifiers and constraints in figure \ref{fig:TelescopicConstraints}. It begins with the $Start$ rule, filling in T and what is known about $\ctxt{}$ and $\mty{}$. Constraint contexts are translated into branches for $\dupCon$ or `situated' constraints that do not mention their context. Instead, their semantics depend on their position in the tree structure: $\btell{x:\mty}$ requires a binding at that exact location and the behaviour of $\bask{x:\mty}$ depends on the bindings in its scope.

\begin{figure}[h]
\begin{tabular}{l c c r c l r}
&&&$\Gamma, \{\exCon{\mty}\}$&$,$&$\llbracket T \rrbracket \; \mty$&$(Start)$\\
$\llbracket x \rrbracket \;$&$r$&$=$&$\{\bask{x:r}\}$&&&$(Var)$\\
$\llbracket \lambda x . T \rrbracket \;$&$r$&$=$&
  $\{\exCon{\mty{p}}, \exCon{\mty{r}}, 
   r \eqCon \mty{p} \;\funcr\; \mty{r}, \btell{x:\mty{p}}\}$
   &$,$&$\llbracket T \rrbracket \; \mty{r}$&$(Lam)$\\
$\llbracket {T\!f} \;\;\; {T\!p} \rrbracket \;$&$r$&$=$&
  $\{ \exCon{\mty{f}},\exCon{\mty{p}}, \mty{p} \;\funcr\; r \eqCon \mty{f}\}$
  &$|$&$\llbracket {T\!f} \rrbracket \; \mty{f}$&$(App)$\\
&&&&$|$&$\llbracket {T\!p} \rrbracket \; \mty{p}$&
\end{tabular}
\caption{STLC Telescopic Tree Semantics}
\label{fig:TelescopicSTLC}
\end{figure}

Given the translation between the two styles of context constraint and branching, the rules in figure \ref{fig:IASTLC} and figure \ref{fig:TelescopicSTLC} can each be derived from the other in a mechanical fashion. Insofar as any typechecker can be explained in terms of the typing rules it can be explained at least as well in terms of the tree.

As an example, here is the tree for $(\lambda x.x) \; y$ in an empty context:


\begin{tabular}{r l}
$\{\},\{\exCon{\mty{}}\}, \{\exCon{\mty{f}}, \; \exCon{\mty{p}}, \; \mty{p} \to \mty \eqCon \mty{f}\}$&$|\{\exCon{\mty{p^\prime}}, \; \exCon{\mty{r}}, \; \mty{f} \eqCon \mty{p^\prime} \to \mty{r}, \; \btell{x:\mty{p^\prime}}\}, \{\bask{x:\mty{r}}\}$\\
&$|\{\bask{y:\mty{p}}\}$
\end{tabular}

With simple types, it is safe to lift the quantifiers to the root of the tree followed by the non-context constraints. Solving the situated context constraints leaves the traditional logical semantics of an unscoped constraint store, followed by context data. A conventional checker can now be recovered by fusing all these processes and solving equality constraints with unification as they arise. 

By making use of the scoped metavariables and additional situated constraints the techniques in Gundry et al.\ \cite{52a673f04e1b4e28aadc29ad0c971b6f} can readily be adapted to implement Hindley-Milner inference as well -- see appendix \ref{app:HM}. This constraint-based formulation also owes much to Bastiaan Heeren\cite{heeren05TopQuality}, while Gundry's thesis\cite{GundryThesis} shows how to extend the techniques to a dependently typed setting.

\section{Related \& Future Work}

Discussion of solving $\dupCon$ at various modes in a linear system recently proved helpful to McBride\cite{1807.04085}.

Elaboration problems are worth exploring: they often involve destroying information due to syntactic sugar, and the reversible nature of information effects' original setting may prove useful as well. Documenting where information is destroyed in one direction also shows where it could be `recreated' from logs to reverse the process. Careful tracking of constraints may also help show that if the elaborator succeeds, so does target typechecking. Edwin Brady and I plan to try this using the core of Idris.

More generally I would like to explore using telescopic constraint trees to help in debugging type errors, distinguishing where constraints are unsolveable due to lack of information from where they are known unsatisfiable. Ultimately I hope to help users trace the flow of failure to its source(s).

\section{Conclusion}

While the type systems presented here are simple and familiar, the presentation itself offers real advantages. The linear-by-default discipline enforced by information effects allows the information aware style to shed light on questions about data and information flow.

The ease with which telescopic constraint trees are derived from information aware systems is backed by -- even at this informal level -- a high degree of confidence that they are a truly general representation of typechecking problems. Put simply, they retain the information in an information aware typing. This can then be used to calculate a typechecker with desired properties.

\newpage

\bibliographystyle{eptcs}
\bibliography{references}

\appendix
\section{IASTLC Modes}
\label{app:modes}
Here are some examples of modes the rules in figure \ref{fig:IASTLC} can be used in, and common ways of describing them. The `Unidirectional' column covers descriptions that ignore the difference between checking and synthesis, while the `Bidirectional' column adds that detail.

{
\renewcommand{\arraystretch}{1.5}
\setlength{\tabcolsep}{1em}

\begin{tabular}{l l l}
Mode&Unidirectional&Bidirectional\\
\hline
$\ctxt^+ \turnstile T^+ : \mty^+$&Type Checking&Checking\\
$\ctxt^+ \turnstile T^+ : \mty^-$&&Synthesis, Inference\\
$\ctxt^- \turnstile T^+ : \mty^+$&Free Variable Analysis&with checked types\\
$\ctxt^- \turnstile T^+ : \mty^-$&&with synthesised types\\
$\ctxt^+ \turnstile T^- : \mty^+$&Proof Search, Program Synthesis&\\
&&\\
\end{tabular}
}

\newpage

\section{Annotations}
\label{app:annotation}

\begin{wrapfigure}{r}{0pt}
$
\inferrule{\mty{a} \dupCon^{\mty{ap}}_{\mty{af}}\\\\
\ctxt{f} \assignCon \ctxt \; \consCon x : \mty{ap}\\\\
\ctxt{f} \turnstile T : \mty{r}\\\\
\mty{f} \eqCon \mty{af} \funcr \mty{r}}
{\ctxt \turnstile \lambda x : \mty{a} \; . \; T \; : \; \mty{f}}
{ALam}$
\label{fig:ALamRule}
\end{wrapfigure}

Annotations are redundant in STLC, and it requires type duplications to handle them. The preferable approach is to duplicate the annotation, pushing one copy in to `check' and returning the other as if `synthesised'.

There are alternative ways to write the rule. Sadly information effects do not go so far as to force the reinvention of bidirectional typechecking. The only requirement is that the annotation is consistent with both the context of the lambda's body and the resulting function type. This could also be achieved by modifying $Lam$ to duplicate $\mty{p}$ or even $\mty{f}$ and check that against the annotation, making it less clear that a partial result can be returned immediately.

\begin{figure}[h]
\begin{tabular}{l c c r c l r}
$\llbracket \lambda x : \mty{a} \; . \; T \rrbracket \;$&$r$&$=$&
  $\{\exCon{\mty{ap}}, \exCon{\mty{af}}, \mty{a} \dupCon^{\mty{ap}}_{\mty{af}}, 
    \btell{x:\mty{ap}}, r \eqCon \mty{af} \funcr \mty{r}\}$
   &$,$&$\llbracket T \rrbracket \; \mty{r}$&$(ALam)$\\
\end{tabular}
\caption{Telescopic Constraint Tree semantics for Annotated Lambda}
\end{figure}

\section{Hindley-Milner}
\label{app:HM}

Figure \ref{fig:IAHM} contains typing rules for a system in the vein of Hindley-Milner, while figure \ref{fig:TelescopicHM} gives a corresponding telescopic constraint tree semantics (use $StartM$ for monomorphic results, $StartP$ for polymorphic). Both use the constraints in figure \ref{fig:HMConstraints}.

For consistency, this presentation uses $n$-ary polytypes $\pty{} ::= \forall \bar{a} . \mty{}$. For $n=0$, we have $\forall.\mty{}$ isomorphic to $\mty{}$, leaving the $App$ and $Lam$ rules isomorphic to STLC's -- the hallmark of rank-1 polymorphism. All bindings are of the form $x:\pty{}$. This means the context doesn't tell us which variables are let-bound, but simplifies instantiation.

Where as Gundry et al.\cite{52a673f04e1b4e28aadc29ad0c971b6f} use $\fatsemi$ as a separator, I have chosen to pair $\genConO$ with $\genConCshort$. During solving, the two delimit the quantifiers to be generalised and $\genConCshort$ does the actual generalisation. I have also written $\ctxt{}\dupCon\{\bar{\ctxt{}}\}$ as sugar for $n$-ary context duplication.

\begin{figure}[h]
\begin{tabular}{r|l l}
$\exCon{\pty{}}$&Bind an unknown $\pty{}$\\
$\exConIs{\pty{}}\mty$&Bind $\pty{}$ with current solution\\
$x:\pty{} \inCon \ctxt$&Binding in context\\
$\ctxt \assignCon \ctxt \;\consCon x:\pty{}$&Context extension\\
$\bask{x:\pty}$&Query/Ask for current binding [here]\\
$\btell{x:\pty}$&Generate/Tell about binding [here]\\
$\pty{} \instCon \mty{}$&$\mty{}$ is an instance of $\pty{}$\\
$\pty{} \genCtxtCon{\mty{}}{\ctxt{}}$&$\pty{}$ generalises $\mty{}$ in $\ctxt{}$\\
$\genConO$&Begin generalisation region/`separator'\\
$\genConC{\pty{}}{\mty{}}$&End region/generalise $\mty{}$ into $\pty{}$
\end{tabular}
\caption{Hindley-Milner constraints}
\label{fig:HMConstraints}
\end{figure}

\begin{figure}[h]
\begin{mathpar}
\inferrule*{\ctxt{f} \assignCon \ctxt{} \; \consCon x : \forall.\mty{p}\\\\
\ctxt{f} \turnstile T : \mty{r}\\\\
\mty{f} \eqCon \mty{p} \;\funcr\; \mty{r}}
{\ctxt{} \turnstile \lambda x . T \; : \; \mty{f}}{Lam}

\inferrule*{\ctxt{}\dupCon^{\ctxt{f}}_{\ctxt{p}}\\\\
\ctxt{f} \turnstile {T\!f} : \mty{f}\\
\ctxt{p} \turnstile {T\!p} : \mty{p}\\\\
\mty{p} \;\funcr\; \mty{r} \eqCon \mty{f}}
{\ctxt{} \turnstile {T\!f} \;\;\; {T\!p} \;:\; \mty{r}}
{App}

\\

\inferrule*{x:\pty \inCon \ctxt\\\\
\pty \instCon \mty}
{\ctxt \turnstile x \; : \; \mty}{Var}

\inferrule*{\ctxt{}\dupCon\{\ctxt{B},\ctxt{gen},\ctxt{T}\} \\\\
\ctxt{B} \turnstile Tb : \mty{b}\\\\
\pty{} \genCtxtCon{\mty{b}}{\ctxt{gen}}\\\\
\ctxt{T^\prime} \assignCon \ctxt{T} \; \consCon x : \pty{}\\\\
\ctxt{T^\prime} \turnstile Tt : \mty{}
}
{\ctxt{} \turnstile \textrm{let}\; x \; = \; Tb \;\textrm{in}\; Tt \;:\; \mty{}}
{Let}
\end{mathpar}
\caption{Information Aware Hindley-Milner rules}
\label{fig:IAHM}
\end{figure}

\begin{figure}[h]
\begin{tabular}{l c c r c l r}
&&&$\Gamma, \{\exCon{\mty}\}$&$,$&$\llbracket T \rrbracket \; \mty$&$(StartM)$\\
&&&$\Gamma, \{\exCon{\pty}, \genConO{},\exCon{\mty}, \genConC{\pty}{\mty}\}$&$,$&$\llbracket T \rrbracket \; \mty$&$(StartP)$\\
$\llbracket x \rrbracket \;$&$r$&$=$&$\{\exCon{\pty},\bask{x:\pty},\pty \instCon r\}$&&&$(Var)$\\
$\llbracket \lambda x . T \rrbracket \;$&$r$&$=$&
  $\{\exCon{\mty{p}}, \exCon{\mty{r}}, 
   r \eqCon \mty{p} \;\funcr\; \mty{r}, \btell{x:\mty{p}}\}$
   &$,$&$\llbracket T \rrbracket \; \mty{r}$&$(Lam)$\\
$\llbracket {Tf} \;\;\; {Tp} \rrbracket \;$&$r$&$=$&
  $\{ \exCon{\mty{f}},\exCon{\mty{p}}, \mty{p} \;\funcr\; r \eqCon \mty{f}\}$
  &$|$&$\llbracket {Tf} \rrbracket \; \mty{f}$&$(App)$\\
&&&&$|$&$\llbracket {Tp} \rrbracket \; \mty{p}$&\\
\end{tabular}
\begin{tabular}{l c c r c l r}
$\llbracket \textrm{let}\; x \; = \; Tb \;\textrm{in}\; Tt \rrbracket \;$&$r$&$=$&
  $\{\exCon{\pty} \}$
  &$|$&$\{\genConO{},\exCon{\mty}, \genConC{\pty}{\mty} \} , \llbracket {Tb} \rrbracket \; \mty$&$(Let)$\\
&&&&$|$&$\{\btell{x:\pty}\} , \llbracket {Tt} \rrbracket \; \mty{r}$&
\end{tabular}
\caption{Hindley-Milner Telescopic Tree Semantics}
\label{fig:TelescopicHM}
\end{figure}






\end{document}